\documentclass[twocolumn]{article}

\usepackage[T1]{fontenc}

\usepackage{geometry}
\usepackage{setspace}
\usepackage{graphicx}
\usepackage{float}
\usepackage{natbib}
\usepackage[version = 4]{mhchem} 
\usepackage{graphicx}
\usepackage{dcolumn}
\usepackage{bm}
\usepackage{array}
\usepackage{float}
\usepackage{mathrsfs}
\usepackage{multirow}
\usepackage{varwidth}
\usepackage{amsmath}
\usepackage{amssymb}
\usepackage{stmaryrd}
\usepackage{graphicx}
\usepackage{hyperref}
\usepackage[switch]{lineno}

\usepackage{subcaption}
\usepackage{cancel}
\usepackage{algorithm}
\usepackage{algorithmic}
\usepackage{booktabs}
\usepackage{makecell}
\usepackage{amsthm}
\usepackage{xcolor}

\usepackage{caption}

\usepackage{authblk}
\author[1]{Runhong He}
\affil[1]{Key Laboratory of System Software (Chinese Academy of Sciences), Institute of Software, Chinese Academy of Sciences, Beijing 100190, China.}
\author[2]{Chao Liu}
\affil[2]{Shanghai Key Laboratory of Trustworthy Computing, East China Normal University, Shanghai 200062, China.}
\author[1]{Xin Hong}
\author[1]{Qiaozhen Chai}
\author[3]{Junyuan Zhou}
\affil[3]{MindSpore Quantum Special Interest Group, ShenZhen 518048, China}
\author[1]{Ji Guan}
\author[4]{Guolong Cui}
\affil[4]{Arclight Quantum Co., LTD. Chinese Academy of Sciences, Beijing 101408, China.}
\author[1]{Shenggang Ying*}

\title{Hamiltonian-Aware ADAPT Variational Quantum Eigensolver for Molecular Ground-State Simulation}

\date{*Email: yingsg@ios.ac.cn}

\begin{document}
\maketitle
\begin{abstract}

Designing compact ansätze in Variational Quantum Eigensolver (VQE) is crucial for solving energetic problems of practical molecules on near-term quantum devices. However, existing Adaptive Derivative-Assembled Pseudo-Trotter (ADAPT) ansätze face two challenges: improper operator selection and accumulation of degraded operators.
In this paper, we propose the Hamiltonian-Aware (HA) ADAPT-VQE algorithm to address these issues.
First, we establish a novel excitation operator selection criterion. It breaks the local constraint of existing criteria by incorporating Hamiltonian information, prioritizes physically meaningful excitation operators, and incurs no extra classical or quantum computational overhead.
Furthermore, we develop a problem-adaptive method for discriminating and pruning redundant excitation operators stemming from improper selection and inevitable degradation. This method balances redundant operator pruning and convergence guarantee, and is applicable to ansätze with arbitrary scales.
Systematic numerical experiments on typical strongly correlated molecular systems demonstrate that our HA-ADAPT-VQE avoids  energy plateaus and outperforms baseline algorithms in terms of energy error, ansatz size, and measurement cost. This work offers an efficient, robust ansatz construction paradigm, facilitating the development and practical deployment of large-scale VQE in quantum chemistry.

\end{abstract}
\maketitle

\section{Introduction}\label{sec:Introduction}
The calculation of molecular energy levels is a fundamental task in
quantum chemistry, as it underpins the understanding of molecular
stability, reactivity, and electronic properties --- key factors
that govern the design and development of new materials, catalysts,
and pharmaceuticals \cite{molecular_property_calculations, Eyring_equation}.
However, classical exact methods, such as the Full Configuration Interaction
(FCI) method \cite{szabo}, are only applicable to the smallest molecular
systems with a handful of electrons, as their computational complexity
scales exponentially with the system size and quickly becomes computationally
intractable for larger molecules. Some approximate approaches, for
instance the Coupled-Cluster Singles and Doubles (CCSD) method \cite{CCSD}
and density functional theory \cite{DFT,DFT_1}, are developed
to balance accuracy and computational cost. However, they frequently
produce qualitative errors when applied to molecules with strong electron
correlations \cite{strongly_correlated_systems,CCSD_fail_1,CCSD_fail_2}.

Quantum computing \cite{qc_nielsen} has emerged as a transformative
paradigm to overcome the inherent limitations of classical computational
methods for such challenging problems. The Quantum Phase Estimation
(QPE) \cite{QPE} was the first quantum algorithm capable in theory
of high-precision energy calculations with polynomial computational
cost. However, the quantum resources demanded by QPE far exceed the
hardware capability of current Noisy Intermediate-Scale Quantum (NISQ)
devices \cite{NISQ,NISQ_1} that are constrained by limited qubit
numbers, short coherence lifetimes, and non-negligible quantum decoherence
and noise.

In comparison, the Variational Quantum Eigensolver (VQE) \cite{VQE_first,VQE_Science,VQE_review_0,VQE_review_1,VQE_review_2}
leverages a hybrid quantum-classical framework and is well-suited
for NISQ devices. In VQE, the parameterized quantum circuit (ansatz) serves as a quantum subroutine for preparing trial states and measuring system energy.
Meanwhile, a classical optimizer iteratively adjusts
the variational parameters to drive convergence to the molecular ground-state
energy. Inspired by the well-established CCSD, the Unitary Coupled
Cluster Singles and Doubles (UCCSD) ansatz \cite{VQE_ucc_review}
has become a popular choice for VQE due to its clear physical and
chemical interpretability. Moreover, compared with hardware-efficient
ansätze \cite{VQE_HEA_nature}, the UCCSD ansatz is more robust against
the barren plateau problem \cite{Barren_plateaus}, i.e., the exponential
vanishing of gradients as the number of qubits increases, which severely
impedes parameter optimization.

Despite these merits, the size of the UCCSD ansatz scales quartically
with the number of electrons, precluding its application to large
molecular systems. Unlike UCCSD, which fixes a problem-agnostic ansatz
upfront, the Adaptive Derivative-Assembled Pseudo-Trotter (ADAPT)
VQE \cite{VQE_ADAPT,VQE_ADAPT_application,VQE_ADAPT_measurement_reuse,VQE_ADAPT_pool_simutameously_measurement,VQE_ADAPT_qubit,VQE_ADAPT_TETRIS,VQE_ADAPT_amp_reording_batch,VQE_ADAPT_CEO,VQE_ADAPT_reduced_RDM}
constructs a compact, problem-tailored ansatz iteratively. In each
iteration, it adds the excitation operator that most effectively contributes
to reducing the variational energy to the existing ansatz. ADAPT-VQE evaluates the competitiveness of excitation operators by computing their gradients at zero parameter values. However, this gradient-based selection criterion is not infallible. Since these gradients are strictly local quantities, a larger gradient does not necessarily guarantee a more favorable energy contribution after parameter optimization.

Recently, Param-ADAPT-VQE \cite{VQE_ADAPT_Param} utilizes operator parameter magnitudes from local VQE optimization as its core screening criterion. This modification provides
a more informative metric for operator selection and avoids the surge
in measurement costs observed in energy-based strategies that rely
on global optimization \cite{VQE_ADAPT_QEB,VQE_ADAPT_ES}. Nevertheless,
in some strongly correlated systems, such as water molecules with
stretched bonds, Param-ADAPT-VQE still suffers a non-negligible energy
plateau, where the system energy ceases to decrease despite continued
operator additions, indicating that the algorithm fails to select
the critical excitation operators. This issue stems from the inherent
locality of both gradient- and parameter-based criteria, which only
focus on the candidate operator. Thus, incorporating nonlocal
information into the excitation-operator selection while avoiding
excessive growth of measurement cost constitutes a promising optimization
direction.

Another challenge of ADAPT-VQE is the inevitable accumulation of degraded
operators. As iterations proceed, the contributions of some operators
added in early steps may be taken over by newly appended ones, rendering
the former redundant even though they were optimal at the beginning.
This is known as the ``fading'' and ``reordering''
of operators \cite{VQE_ADAPT_Pruned}. Pruned-ADAPT-VQE \cite{VQE_ADAPT_Pruned} evaluates the redundancy of each operator based on its parameter magnitude and position within the ansatz. Typically, operators with small parameter values that appear at early positions are preferentially removed. Nevertheless, its combination of inverse-square weighting for parameters and exponential weighting for positions cannot effectively eliminate redundant operators in the middle and late stages of the ansatz. As the ansatz grows longer, the exponential positional terms rapidly overwhelm the polynomial parameter weights, which degrades the pruning performance.

To overcome the aforementioned limitations, we present an enhanced ADAPT-VQE variant. First, during operator selection, we integrate locally optimized parameters with corresponding Hamiltonian coefficients to incorporate nonlocal information. This is partially
inspired by the success of Hamiltonian-Informed (Hi) \cite{VQE_HiUCCSD}
UCCSD, which filters out poor operators based on their vanishing Hamiltonian
coefficients. Our method breaks the locality constraint of conventional ADAPT-VQE while introducing no extra classical and quantum computational cost. We term this new algorithm Hamiltonian-Aware (HA) ADAPT-VQE. Second, we develop a problem-adaptive pruning scheme equipped with a dynamically adjustable tolerance, enabling robust pruning for ansätze of variable scales.
Numerical benchmarks on typical molecular systems demonstrate that
HA-ADAPT-VQE avoids stagnant energy plateaus and outperforms the
baseline approaches (standard ADAPT-VQE, Param-ADAPT-VQE, and Pruned-ADAPT-VQE)
in energy error, ansatz size and measurement overhead.

The remainder of this paper is structured as follows: Sections \ref{subsec:2.1} and \ref{subsec:2.2} briefly introduce VQE, ADAPT-VQE and Param-ADAPT-VQE. Section \ref{subsec:2.3} details the proposed HA-ADAPT-VQE algorithm. Section \ref{sec:3} presents numerical experiments on typical strongly correlated molecular systems and compares the performance of our method with baseline approaches. Finally, Section \ref{sec:4} concludes this work.

\section{Methods }\label{sec:2}

\subsection{VQE Algorithm}\label{subsec:2.1}

In accordance with conventional notation \cite{VQE_ucc_review,VQE_ADAPT,VQE_first},
we use $i,j,k,...$ to label occupied spin orbitals, $a,b,c,...$
for virtual (unoccupied) spin orbitals, and $p,q,r,s$ for general spin
orbitals irrespective of occupation. The total number of spin orbitals
is denoted as $N$.

Under the Born-Oppenheimer approximation, which neglects the effects
of nuclear motion, the molecular Hamiltonian in second-quantization
form reads: \cite{szabo}
\begin{equation}
\hat{H}=\sum_{p,q}^{N}h_{q}^{p}\hat{a}_{p}^{\dagger}\hat{a}_{q}+\frac{1}{2}\sum_{p,q,r,s}^{N}h_{rs}^{pq}\hat{a}_{p}^{\dagger}\hat{a}_{q}^{\dagger}\hat{a}_{r}\hat{a}_{s},
\end{equation}
where the creation operator $\hat{a}_{i}^{\dagger}$ and annihilation
operator $\hat{a}_{i}$ obey the fermionic anti-commutation relations:
\begin{equation}
\left\{ \hat{a}_{i},\hat{a}_{j}^{\dagger}\right\} =\delta_{i,j},\quad\left\{ \hat{a}_{i},\hat{a}_{j}\right\} =\left\{ \hat{a}_{i}^{\dagger},\hat{a}_{j}^{\dagger}\right\} =0.
\end{equation}
The one- and two-electron integrals $h_{q}^{p}$ and $h_{rs}^{pq}$
can be computed by classical computers within a specified basis set
(e.g., the STO-3G basis set) \cite{szabo}. 

According to the variational principle \cite{RRvariational}, the
expectation value of the Hamiltonian with respect to a trial state
$|\Psi(\vec{\theta})\rangle$ provides an upper bound for the unknown
ground-state energy $E_{0}$, i.e., $\langle\Psi(\vec{\theta})|H|\Psi(\vec{\theta})\rangle\geq E_{0}$.
In VQE, the trial state is prepared by an ansatz $U(\vec{\theta})$
with moderate size from an initial reference state $|\Psi_{0}\rangle$,
namely, $|\Psi(\vec{\theta})\rangle=U(\vec{\theta})|\Psi_{0}\rangle$.

The UCCSD ansatz takes the form $U(\vec{\theta})=\text{e}^{\hat{T}(\vec{\theta})}$,
with \cite{VQE_ucc_review}
\begin{equation}
\hat{T}(\vec{\theta})=\hat{T}_{1}(\vec{\theta})+\hat{T}_{2}(\vec{\theta}),
\end{equation}
where the single excitation operators are defined as 
\begin{equation}
\begin{alignedat}{1}\hat{T}_{1}(\vec{\theta}) & =\sum_{i,a}\theta_{i}^{a}\hat{\tau}_{i}^{a}=\sum_{i,a}\theta_{i}^{a}(\hat{t}_{i}^{a}-\hat{t}_{a}^{i})\\
 & =\sum_{i,a}\theta_{i}^{a}(\hat{a}_{a}^{\dagger}\hat{a}_{i}-\hat{a}_{i}^{\dagger}\hat{a}_{a}),
\end{alignedat}
\end{equation}
and the double excitation operators are 
\begin{equation}
\begin{alignedat}{1}\hat{T}_{2}(\vec{\theta}) & =\sum_{a>b,i>j}\theta_{ij}^{ab}\hat{\tau}_{ij}^{ab}=\sum_{a>b,i>j}\theta_{ij}^{ab}(\hat{t}_{ij}^{ab}-\hat{t}_{ab}^{ij})\\
 & =\sum_{a>b,i>j}\theta_{ij}^{ab}(\hat{a}_{a}^{\dagger}\hat{a}_{b}^{\dagger}\hat{a}_{i}\hat{a}_{j}-\hat{a}_{i}^{\dagger}\hat{a}_{j}^{\dagger}\hat{a}_{a}\hat{a}_{b}).
\end{alignedat}
\end{equation}

For practical implementation on NISQ devices, the first-order Trotter-Suzuki
approximation \cite{VQE_trotter_one} can be employed to simplify the exponential of non-commuting
operators, achieving a balance between accuracy and circuit feasibility:
\begin{equation}
\text{e}^{\hat{T}(\vec{\theta})}\thickapprox\prod_{i,a}\text{e}^{\theta_{i}^{a}\hat{\tau}_{i}^{a}}\cdot\!\!\!\prod_{i>j,a>b}\text{e}^{\theta_{ij}^{ab}\hat{\tau}_{ij}^{ab}}.
\end{equation}
For the basic component $\text{e}^{\theta_{i}\hat{\tau}_{i}}$, popular
circuit implementations include Pauli string simulation \cite{VQE_ADAPT_qubit,Jordan_Wigner}
and efficient circuit realization \cite{VQE_ADAPT_CEO,VQE_fops_9_cnot,VQE_fops_13_cnot}.

\subsection{ADAPT-VQE and Param-ADAPT-VQE }\label{subsec:2.2}

As a fixed and problem-agnostic ansatz, UCCSD scales as $\mathcal{O}(N^{4})$, making it challenging to handle large molecules on NISQ devices.
It is worth noting that not all excitation operators
incorporated in the ansatz yield significant contributions. In fact,
many of them are redundant and have negligible effects on the system
energy, which leads to unnecessary resource consumption. For example,
the symmetry-reduced UCCSD \cite{VQE_SymUCCSD} and HiUCCSD \cite{VQE_HiUCCSD,VQE_SymUCCSD_failed}
eliminate numerous ineffective operators by exploiting the point group
symmetry of molecules.

Different from the static UCCSD ansatz, ADAPT-VQE constructs a problem-tailored
ansatz by iteratively appending excitation operators that possess
substantial contributions to the system energy. Specifically,
it defines an excitation operator pool based on UCCSD and initializes the ansatz as the identity operator.
In each iteration, ADAPT-VQE first screens the operator pool by calculating the energy gradient for each operator with a zero-valued parameter under the current ansatz. Then, the operator with the largest gradient
is appended to the ansatz, as it tends to bring the most significant
energy decrease. Subsequently, a global VQE optimization is executed
on the new ansatz. To accelerate convergence and avoid the barren
plateau \cite{VQE_ADAPT_vs_bp}, a ``warm-starting''
strategy can be employed, i.e., the initial values of the optimization
are set to the converged results of the preceding iteration, except
for the newly added parameter, which is initialized to zero. This
iterative process of operator selection---appending---VQE optimization
is continued until the norm of the pool's gradient vector is lower
than a predefined tolerance.

However, the operator with the largest gradient does not necessarily yield the largest energy reduction. To improve robustness, Param-ADAPT-VQE performs an independent local VQE optimization on the parameter of each candidate excitation operator based on the current trial state, and selects the operator with the largest converged parameter magnitude. This effectively enhances the compactness of the ansatz.
Meanwhile, by developing a sub-Hamiltonian (with size scaling as $\mathcal{O}(N^3)$) for local optimization, and hot-starting subsequent optimizations by using locally optimized results to initialize new operators rather than zeros in ADAPT-VQE, a lower measurement cost is observed in experiments compared to ADAPT-VQE.

Nevertheless, Param-ADAPT-VQE still
relies solely on local information and suffers from energy plateaus
when applied to some strongly correlated molecules, such as stretched
\ce{H2O}. 

\subsection{HA-ADAPT-VQE }\label{subsec:2.3}

A limitation of Param-ADAPT-VQE is that it omits that excitation operators
with similar locally optimized parameter magnitudes may yield markedly
different energy reductions. One heuristic improvement is to incorporate
Hamiltonian information to achieve a more reliable estimation of their
energetic contributions. Physically, the Hamiltonian coefficient (electron integrals) $h_{i}^{a}$
quantifies the quantum transition coupling strength between occupied
orbital $i$ and virtual orbital $a$. An operator with larger $|h_{i}^{a}|$ typically 
yields a more prominent energy correction.

For the anti-hermite operators $\hat{\tau}$, the relations $\hat{\tau}^{3}=-\hat{\tau}$
and $\hat{\tau}^{4}=-\hat{\tau}^{2}$ hold \cite{VQE_ADAPT_grad_free}.
After Taylor expansion, we have
\begin{equation}
\text{e}^{\theta\hat{\tau}}=\mathbb{I}+(1-\cos(\theta))\hat{\tau}^{2}+\sin(\theta)\hat{\tau}.
\end{equation}
Consider the reduced Hilbert space spanned by $\{|i\rangle,|a\rangle\}$, where $|i\rangle$ ($|a\rangle$) denotes that spin orbital $i$ ($a$) is occupied. 
We then have $\hat{t}_{i}^{a}|i\rangle=|a\rangle$,  $\hat{t}_{a}^{i}|a\rangle=|i\rangle$ and $\hat{t}_{a}^{i}|i\rangle=\hat{t}_{i}^{a}|a\rangle=0$. The evolution of coupled
cluster can be simplified to \cite{MindQuantum} 
\begin{equation}
\text{e}^{\theta\hat{\tau}_{i}^{a}}\left(\begin{array}{c}
c_{i}|i\rangle\\
c_{a}|a\rangle
\end{array}\right)=\left(\begin{array}{rr}
\cos(\theta), & \sin(\theta)\\
-\sin(\theta), & \cos(\theta)
\end{array}\right)\left(\begin{array}{c}
c_{i}|i\rangle\\
c_{a}|a\rangle
\end{array}\right).
\end{equation}
The Hartree-Fock state $|\text{HF}\rangle=|i\rangle$, i.e., $c_{i}=1$, $c_{a}=0$.

Assuming the reference state is $|i\rangle$, i.e., no effective excitation
operator has previously been applied to spin orbitals $i$ and $a$,
the resulting state reads
\begin{equation}
|\psi\rangle=\text{e}^{\theta\hat{\tau}_{i}^{a}}|i\rangle=\cos(\theta)|i\rangle-\sin(\theta)|a\rangle.
\end{equation}
Within the Hamiltonian, the term directly associated with the interest $\hat{\tau}_{i}^{a}$
is $h_{i}^{a}\hat{t}_{i}^{a}$, whose energetic contribution reads
\begin{equation}
\langle\psi|h_{i}^{a}\hat{t}_{i}^{a}|\psi\rangle=-\frac{1}{2}h_{i}^{a}\sin(2\theta).\label{eq:metric}
\end{equation}
We can select excitation operators using the Hamiltonian-aware metric $\left|\frac{1}{2}h_{i}^{a}\sin(2\theta)\right|$,
or equivalently $\left|h_{i}^{a}\sin(2\theta)\right|$, in contrast
to $\left|\theta\right|$ used in Param-ADAPT-VQE. Note that this quantity
does not represent the turely energetic contribution induced by the
excitation operator $\hat{\tau}_i^a$, since other Hamiltonian terms coupled
to $\hat{\tau}_i^a$ are all omitted \cite{VQE_ADAPT_Param}. It is employed
herein for low-cost approximation.

This strategy can also be intuitively interpreted as follows: for excitation
operators with comparable $|\theta_{i}|$, we prefer that bearing
the largest associated Hamiltonian coefficient $|h_{i}|$, since such
operators correspond to physically meaningful excitations and tend
to yield larger energetic contributions. Especially when iterations
converge toward an energy plateau, all locally optimized parameters
shrink to negligible magnitudes and thus fail to furnish effective
selection criteria. By incorporating Hamiltonian information, 
we can identify more promising candidate excitation operators. 
Based on this feature, we name the proposed algorithm Hamiltonian-Aware
(HA) ADAPT-VQE.

Nevertheless, this approximation remains inherently coarse and lacks
rigorous theoretical guarantees, which may prevent the algorithm from
selecting the truly optimal excitation operators in practical iterative
executions. To mitigate this issue, we further propose a problem-adaptive pruning 
strategy to remove redundant operators caused by improper selection 
as well as inevitable ``fading'' and ``reordering'' \cite{VQE_ADAPT_Pruned}, 
so as to maintain a compact ansatz.

\begin{figure}
\centering
\includegraphics[scale=0.95]{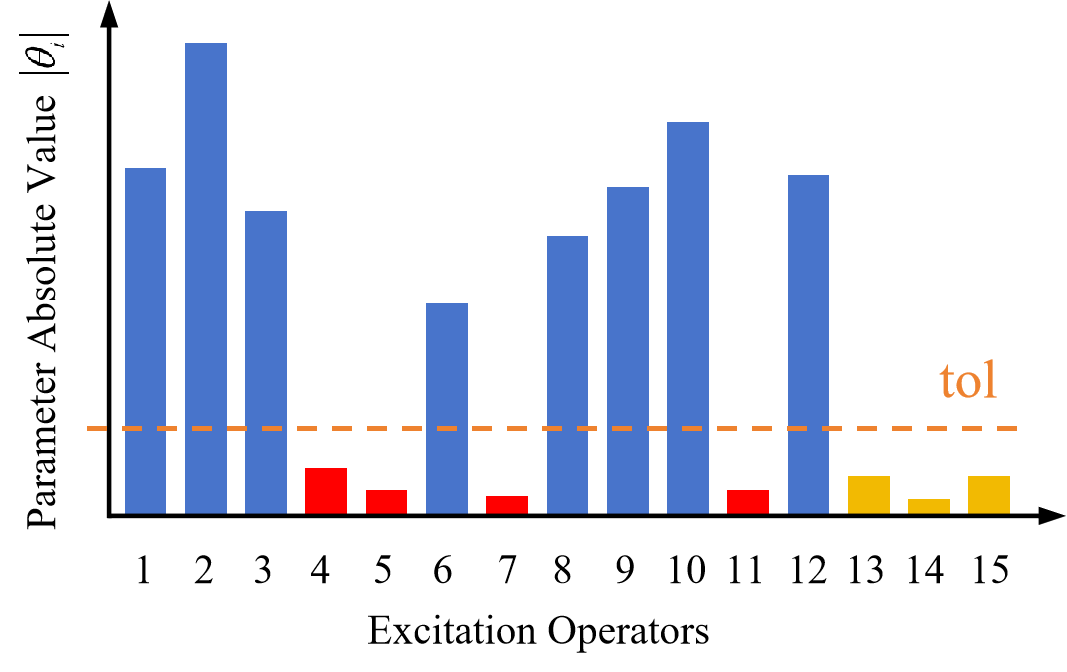}\caption{Excitation operators are divided into three categories according to their parameter magnitudes and positions: (1) Operators with parameter magnitudes exceeding the predefined threshold are non-redundant (marked in blue); (2) Operators below the threshold and located before the last non-redundant operator are redundant candidates and can be removed (marked in red); (3) The remaining operators are tentatively retained to guarantee iterative convergence (marked in yellow). }\label{fig:1}
\end{figure}

Figure~\ref{fig:1} schematically illustrates our proposed problem-adaptive pruning strategy for redundant operators. 
We first initialize a threshold $\mathrm{tol>0}$.
Operators in the ansatz are categorized into three groups according
to their parameters and positions:

1. \textit{Non-redundant operators} (marked in blue): Operators whose parameter magnitudes exceed $\mathrm{tol}$.

2. \textit{Redundant candidate operators} (marked in red): Operators with parameter magnitudes below $\mathrm{tol}$ and located before the last non-redundant operator. These operators are removable.

3. \textit{Temporarily retained operators} (marked in yellow): Operators located after the last non-redundant operator. 
With parameter magnitudes also below  $\mathrm{tol}$, they are kept provisionally to guarantee iterative convergence.

Furthermore, to enhance the generalizability of the proposed strategy,
$\mathrm{tol}$ can configured to adjust adaptively. 
If removing redundant operators leads to a non-negligible energy increase 
in the subsequent global optimization — indicating that important excitation 
operators have been accidentally eliminated — we reduce $\mathrm{tol}$ appropriately
(e.g., by halving its value) to prevent similar improper pruning in later iterations.
This adaptive mechanism ensures stable iterative convergence and renders the algorithm applicable to arbitrary quantum systems. In practice, we may adopt a relatively large initial $\mathrm{tol}$ according to practical scenarios, to preserve only key operators at the early iterations. The algorithm will automatically lower the threshold subsequently to ensure stable operation.

Removing redundant operators from the ansatz can not only mitigate noise effects and thus improve computational accuracy, but also reduce the measurement cost.
The parameter-shift rule \cite{parameter_shift_rule_0, Parameter_shift} allows us to compute the gradient of each variational parameter using the difference in Hamiltonian expectation values between two parameter-shifted quantum circuits. For an ansatz $U(\vec{\theta}^{(m)})$ with $m$ variational parameters, optimization via the Broyden–Fletcher–Goldfarb–Shanno (BFGS) minimizer \cite{bfgs} generally involves $\mathcal{O}(m^2)$ gradient evaluations. Therefore, eliminating redundant operators and reducing the number of parameters leads to a quadratic reduction in measurement cost.

\begin{figure*}
\centering
\includegraphics[width=0.8\paperwidth]{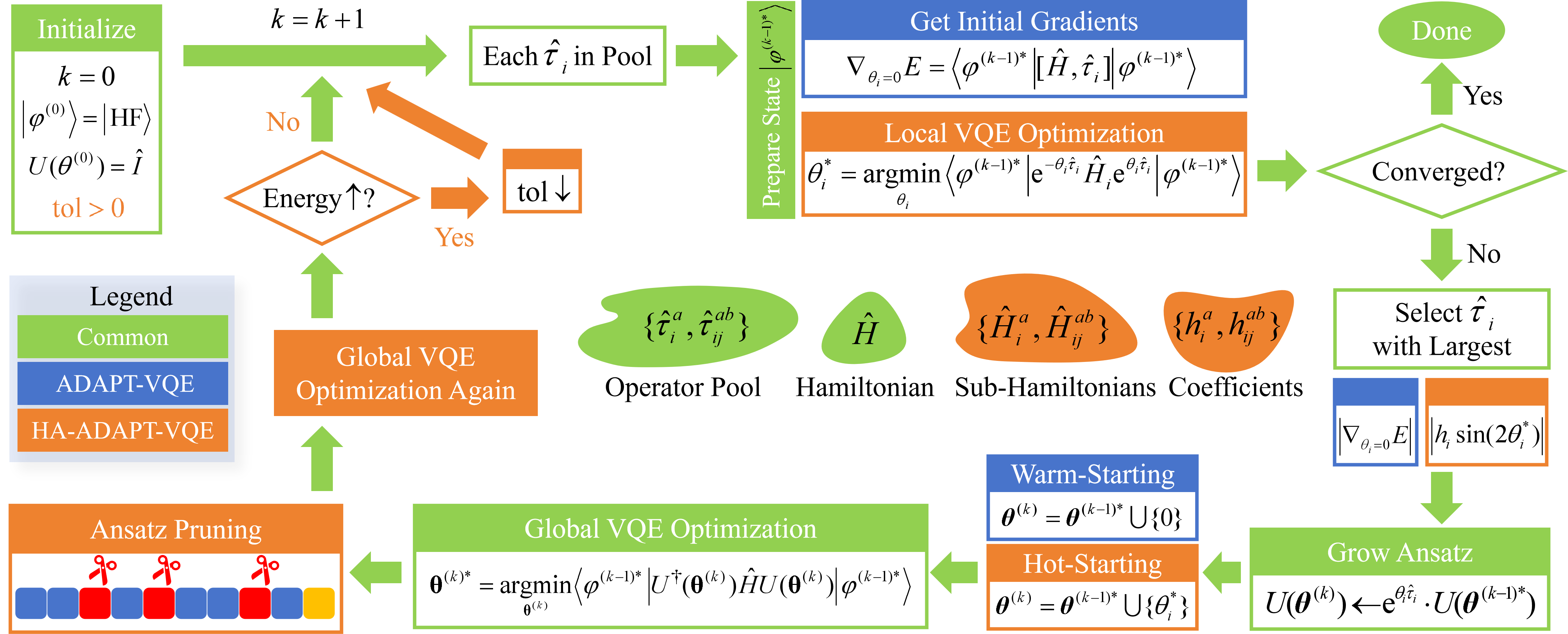}\caption{
Schematic comparison of ADAPT-VQE and the proposed HA-ADAPT-VQE. Unique
modules are highlighted in blue (ADAPT-VQE) and orange (HA-ADAPT-VQE),
whereas shared operations are colored green. Their core differences
lie in (1) the operator selection criterion and (2) the ansatz pruning
step. The sub-Hamiltonian based local VQE optimization and hot-starting
techniques, inherited from Param-ADAPT-VQE, are introduced to reduce
measurement overhead. }\label{fig:2}
\end{figure*}

Next we summarize the workflow of our proposed HA-ADAPT-VQE algorithm. 
Its schematic diagram, along with a comparison against conventional ADAPT-VQE, 
is presented in Figure \ref{fig:2}.

\textbf{Step 1.} Initialization: Define the excitation operator pool
$\mathbb{P}:=\{\hat{\tau}_{i}^{a},\hat{\tau}_{ij}^{ab}\}$. For each
candidate $\hat{\tau}_{i}$ in pool, construct its sub-Hamiltonian
$\hat{H}_{i}$ by extracting those terms from the full Hamiltonian
$\hat{H}$ that sharing indices with $\hat{\tau}_{i}$, namely, $\{p,q,r,s\}\cap\{a,b,i,j\}\neq\emptyset$,
where the two sets correspond to the indices of the Hamiltonian term
$\hat{t}_{i}$ and the excitation operator $\hat{\tau}_{i}$, respectively (i.e., the sub-Hamiltonian technique in Param-ADAPT-VQE).
Collect Hamiltonian
coefficients $h_{i}$, i.e., the electronic integrals of $\hat{t}_{i}$.
Initialize hyperparameters: pruning parameter tolerance $\mathrm{tol}>0$, parameter vector
norm tolerance $\epsilon>0$, and maximum iteration number $k_{\text{max}}\geqslant0$.
Set the initial reference state to the Hartree-Fock state $|\psi^{(0)}\rangle=|\textrm{HF}\rangle$,
iteration counter $k=0$ and initial ansatz $U^{(0)}=\hat{I}$ (identity
operator). 

\textbf{Step 2.} Update the iteration counter via $k\leftarrow k+1$.
If $k>k_{\text{max}}$, terminate the iteration and jump to Step 7;
otherwise commence the $k^{\text{th}}$ iteration as follows: Prepare
the state $|\psi^{(k-1)*}\rangle=U(\vec{\theta}^{(k-1)*})|\psi^{(0)}\rangle$,
where $\vec{\theta}^{(k-1)*}$ denotes the optimized parameters obtained
from the $(k-1)^{\text{th}}$ iteration. We note that in this work, the superscript * refers to the
optimized result, rather than the complex conjugate. 
For every $\hat{\tau}_{i}\in\mathbb{P}$, 
perform local VQE optimization to solve for its optimal parameter
$\theta_{i}^{*}$ via
\begin{equation}
\theta_{i}^{*}=\underset{\theta_{i}}{\text{argmin}}\langle\psi^{(k-1)*}|\text{e}^{-\theta_{i}\hat{\tau}_{i}}\hat{H_{i}}\text{e}^{\theta_{i}\hat{\tau}_{i}}|\psi^{(k-1)*}\rangle.
\end{equation}

\textbf{Step 3.} If the norm of the optimized parameter vector satisfies
$\sqrt{\sum_{i}(\theta_{i}^{*})^{2}}<\epsilon$, terminate iteration
and go to Step 7.

\textbf{Step 4.} Compute the Hamiltonian-aware score $\left|h_{i}\sin(2\theta_{i}^{*})\right|$.
Select the excitation operator yielding the largest score, then prepend
its associated unitary $\text{e}^{\theta_{i}^{*}\hat{\tau}_{i}}$
to the existing ansatz: $U(\vec{\theta}^{(k)})\leftarrow\text{e}^{\theta_{i}^{*}\hat{\tau}_{i}}\cdot U(\vec{\theta}^{(k-1)*})$.
Update the parameters set as $\vec{\theta}^{(k)}\leftarrow\vec{\theta}^{(k-1)*}\cup\{\theta_{i}^{*}\}$ (i.e., the ``hot-starting'' in Param-ADAPT-VQE).

\textbf{Step 5.} Perform global VQE optimization over all variational
parameters in the updated ansatz:
\begin{equation}
\vec{\theta}^{(k)*}=\underset{\vec{\theta}^{(k)}}{\text{argmin}}\langle\psi^{(0)}|U^{\dagger}(\vec{\theta}^{(k)})\hat{H}U(\vec{\theta}^{(k)})|\psi^{(0)}\rangle.
\end{equation}

\textbf{Step 6.} An excitation operator is flagged as redundant if its parameter magnitude is below $\mathrm{tol}$ and subsequent operators have parameter magnitudes exceeding $\mathrm{tol}$.
Prune all redundant operators, then perform global VQE optimization again on the remaining variational parameters. If the system energy increases non-negligibly after pruning, decrease $\mathrm{tol}$ appropriately.

\textbf{Step 7.} Evaluate the full-Hamiltonian expectation value with
respect to the optimized final trial state $|\psi^{(k)*}\rangle=U(\vec{\theta}^{(k)*})|\psi^{(0)}\rangle$,
namely, $\langle\psi^{(k)*}|\hat{H}|\psi^{(k)*}\rangle$, and output it as
the computed molecular ground-state energy.

Note, to further mitigate measurement overhead, we employ two techniques originally proposed in Param-ADAPT-VQE \cite{VQE_ADAPT_Param}: the sub-Hamiltonian scheme and hot-starting initialization. 
The former reduces the number of Hamiltonian terms involved in local optimization
for obtaining $\theta_{i}^{*}$ from $\mathcal{O}(N^{4})$ to $\mathcal{O}(N^{3})$,
while the latter
decreases parameter-update iterations by initializing global VQE optimization
with a starting guess (i.e., $\vec{\theta}^{(k)}\leftarrow\vec{\theta}^{(k-1)*}\cup\{\theta_i^*\}$) closer to the optimum, in contrast to the ADAPT-VQE
warm-starting method $\vec{\theta}^{(k)}\leftarrow\vec{\theta}^{(k-1)*}\cup\{0\}$.

\section{NUMERICAL RESULTS }\label{sec:3}

In this work, we employ PySCF \cite{pyscf} to compute molecular electronic
integrals (Hamiltonian coefficients) under the STO-3G basis set with no frozen orbitals. The
MQChemSimulator of MindSpore Quantum \cite{MindQuantum} is utilized
for ansatz simulation, where each fermionic excitation operator is
treated as an integral operator rather than decomposed into primitive
quantum gates to accelerate simulation. This implementation yields
results equivalent to those obtained via explicit decomposition protocols
documented in refs.~\cite{VQE_fops_13_cnot,VQE_ADAPT_QEB}, where
single and double excitation operators are implemented using 2 and
13 CNOT gates plus several single-qubit gates, respectively. Variational
parameter optimization is performed using the BFGS solver \cite{bfgs} from the SciPy library \cite{scipy}. 
All simulations are performed under ideal conditions, where sampling noise and hardware-related defects are not taken into account. The excitation operator pool is constructed based on HiUCCSD \cite{VQE_HiUCCSD}, which systematically eliminates invalid operators in UCCSD that violate the molecular point group symmetry.
The source code and experimental data supporting the findings
of this study are publicly accessible at
\href{https://atomgit.com/mindspore/mindquantum/tree/research/paper_with_code/HA_ADAPT_VQE}{https://atomgit.com/mindspore/mindquantum/tree} \href{https://atomgit.com/mindspore/mindquantum/tree/research/paper_with_code/HA_ADAPT_VQE}{/research/paper\_with\_code/HA\_ADAPT\_VQE}.

In this section, we benchmark the proposed HA-ADAPT-VQE algorithm
against three baseline approaches (standard ADAPT-VQE, Param-ADAPT-VQE
and Pruned-ADAPT-VQE) on three representative molecules: \ce{BeH2}, \ce{H2O},
and \ce{NH3}, whose equilibrium bond lengths are 1.32 $\text{\AA}$, 1.02 $\text{\AA}$
and 1.09 $\text{\AA}$, respectively.
These molecules are uniformly stretched away from their equilibrium bond lengths to produce substantial correlation energies, so as to evaluate algorithm performance for strongly correlated molecular systems. Specifically,
 $R$(Be-H) = 2.25 $\text{\AA}$
for \ce{BeH2},  $R$(O-H) = 2.40 $\text{\AA}$ for \ce{H2O}, and $R$(N-H) = 2.40 $\text{\AA}$
for \ce{NH3}. The
threshold values for parameter and gradient norm $\epsilon$ are set to $10^{-4}$,
and the maximum allowed ansatz size is capped at 120. We emphasize
that these thresholds are empirically configured to make plotting
ranges comparable among all algorithms for direct performance comparison.
Each algorithm can achieve higher accuracy via stricter threshold
configurations, while the overall curve shapes remain unchanged. The
energy increment tolerance $\mathrm{tol}$ of HA-ADAPT-VQE is initialized
to $5\times10^{-4}$. If the energy increases by more than $1\times10^{-7}$
$\mathrm{Hartree}$ after pruning redundant operators, $\mathrm{tol}$
is halved.

We first compare the overall performance of different algorithms in
terms of energy error, ansatz size and measurement cost from a macroscopic
perspective. Then, we take \ce{H2O} as an example for microscopic 
analysis to explore the underlying mechanisms behind the improved performance of HA-ADAPT-VQE.
\begin{figure*}[htbp]
\centering
\subfloat[]{\includegraphics[width=0.255\paperwidth]{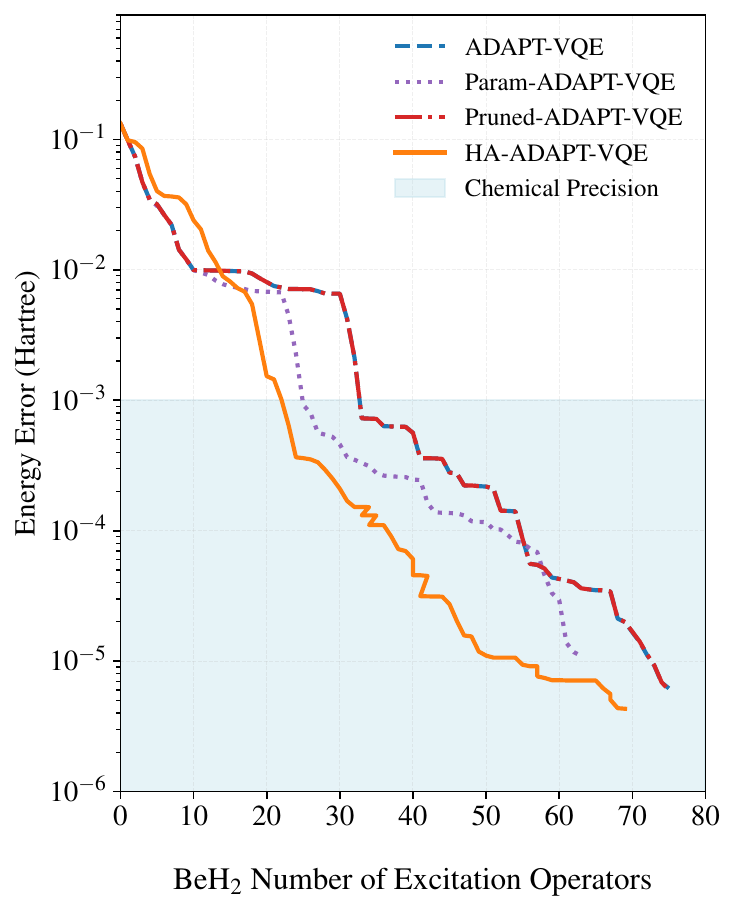}}
\subfloat[]{\includegraphics[width=0.255\paperwidth]{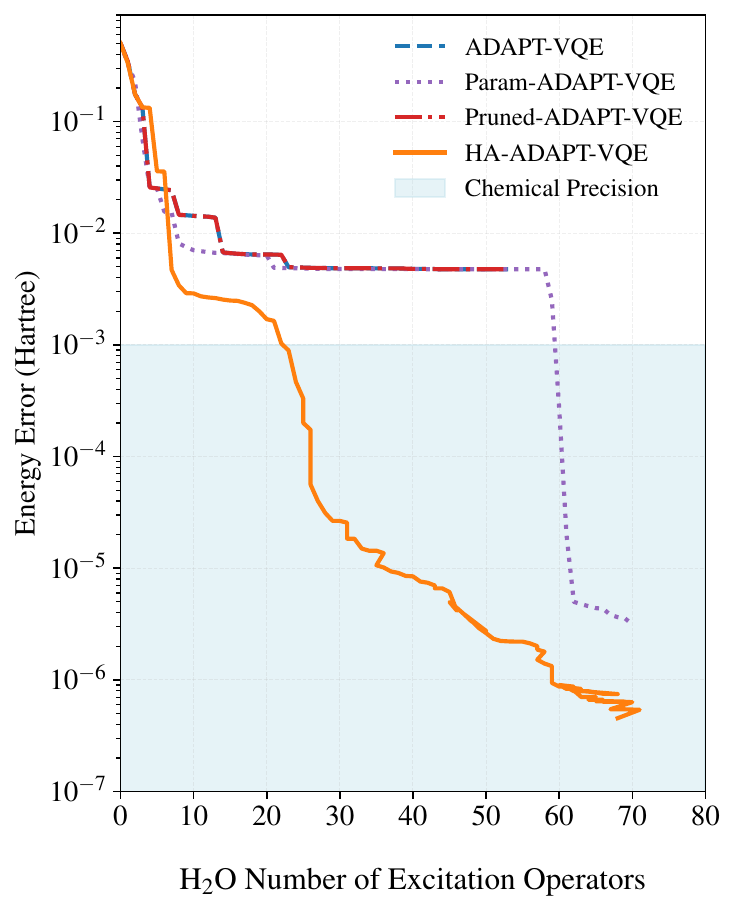}}
\subfloat[]{\includegraphics[width=0.255\paperwidth]{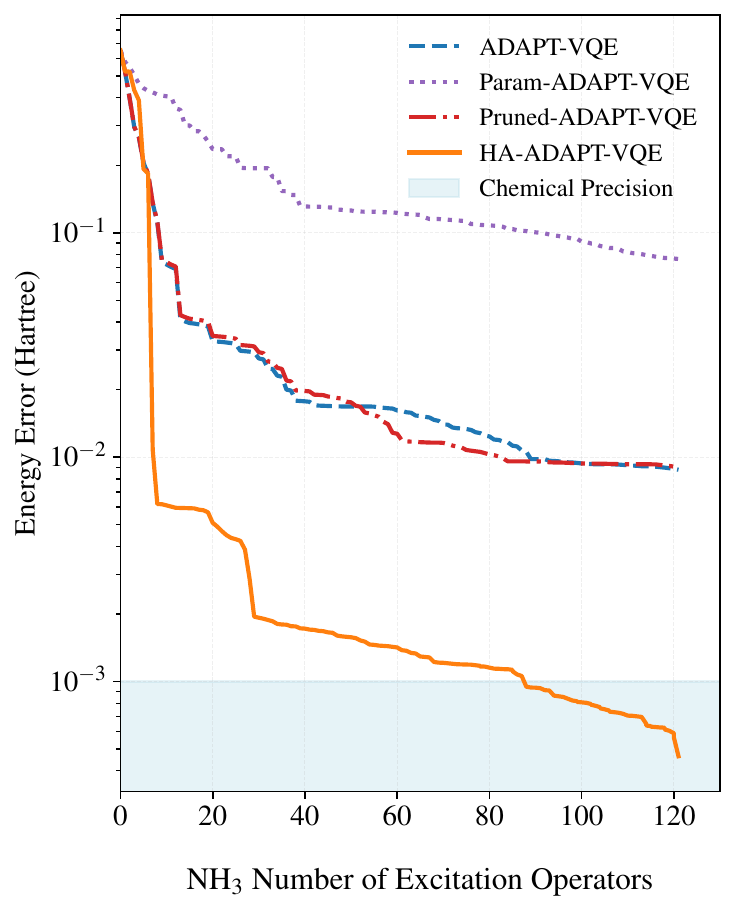}}

\subfloat[]{\includegraphics[width=0.255\paperwidth]{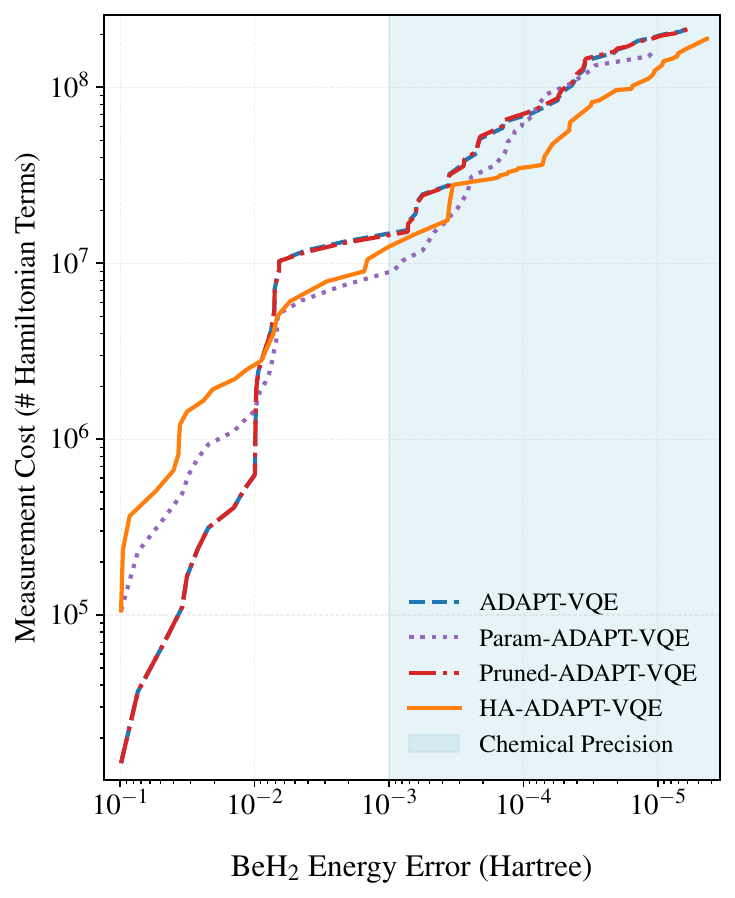}}
\subfloat[]{\includegraphics[width=0.255\paperwidth]{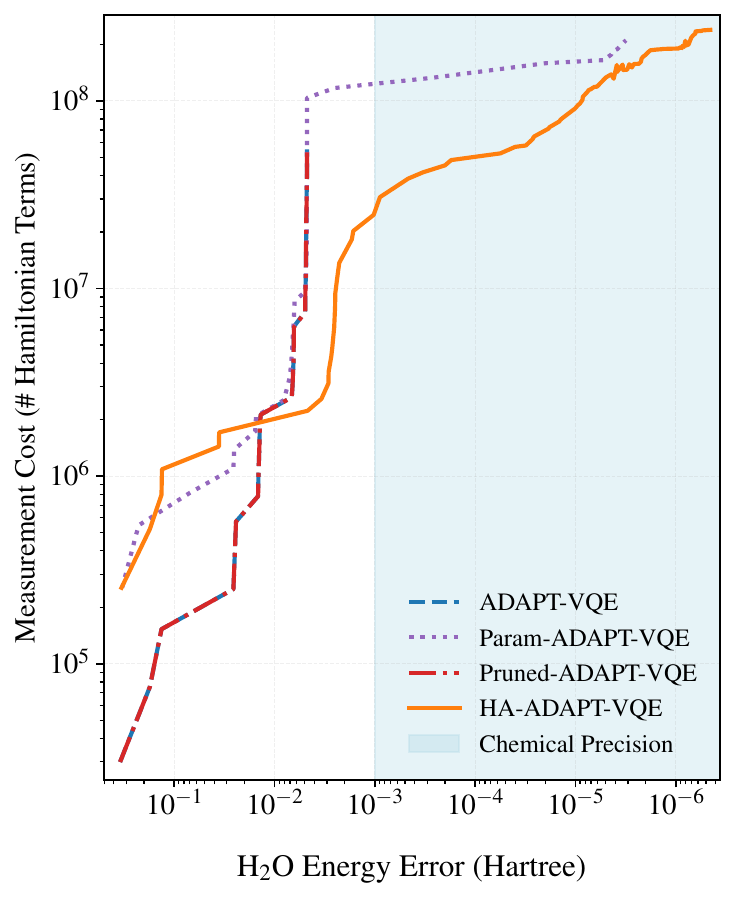}}
\subfloat[]{\includegraphics[width=0.255\paperwidth]{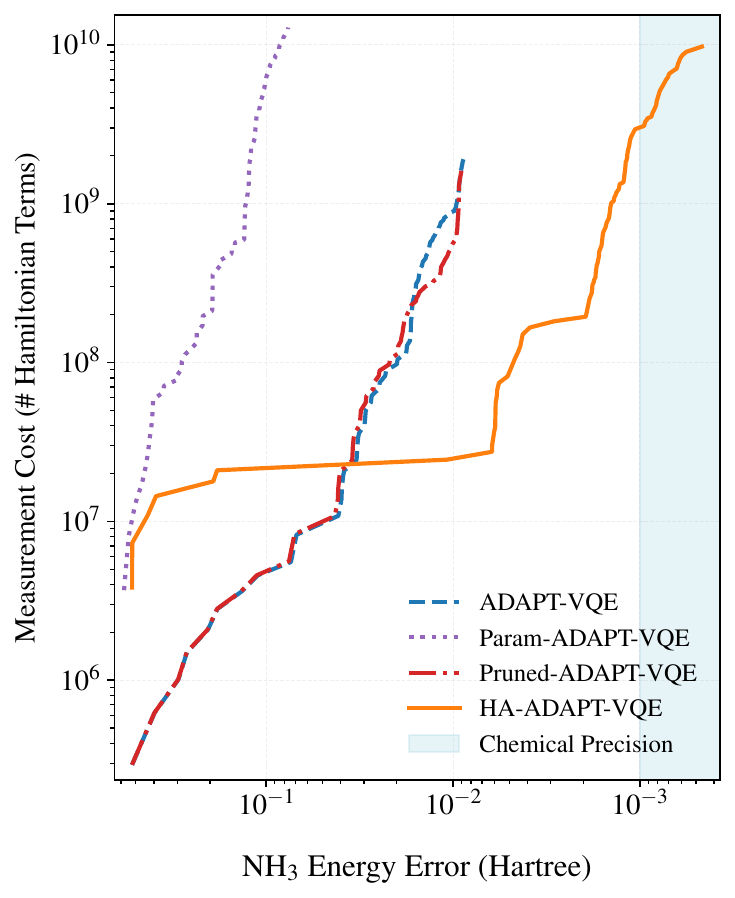}}
\caption{(a\textminus c) Evolution of energy error with iteration count; (d\textminus f)
corresponding measurement cost as a function of energy error for \ce{BeH2},
\ce{H2O}, and \ce{NH3} at uniformly stretched bond lengths of 2.25, 2.40, and
2.40 $\text{\protect\AA}$, respectively. }\label{fig:3}
\end{figure*}

To evaluate the results, we consider the absolute error between the
VQE energy and the exact FCI energy. In addition, to facilitate comparisons
with results from other references, we shade in blue the region where
the error falls below the desirable chemical precision of $10^{-3}$
Hartree \cite{VQE_review_1}. 

Figure~\ref{fig:3} (a)\textminus (c) plot the evolution of energy error
against ansatz size (namely the number of excitation operators) for
the four algorithms on \ce{BeH2}, \ce{H2O} and \ce{NH3}, respectively. Overall, HA-ADAPT-VQE
achieves superior accuracy with a more compact ansatz compared with
competing algorithms. Notably, the curve of HA-ADAPT-VQE sometimes
exhibits \rotatebox[origin=c]{140}{Z}-shaped segments, stemming from the removal of multiple
excitation operators and the consequent reduction in ansatz size.
Such behavior never emerges for Pruned-ADAPT-VQE, as it eliminates
at most one operator per iteration and yields a non-decreasing ansatz
size.
It can be observed that, in most cases, the removal of multiple excitation operators does not cause a significant energy rise, demonstrating that the proposed pruning strategy works well.

Specifically, Figure~\ref{fig:3}(a) illustrates that for \ce{BeH2}, HA-ADAPT-VQE,
ADAPT-VQE, Pruned-ADAPT-VQE and Param-ADAPT-VQE require 23, 33, 33,
and 25 excitation operators, respectively, to reach the chemical precision.
In comparison with the latter three approaches, HA-ADAPT-VQE achieves ansatz size reductions of $30.3\%$, $30.3\%$ and $8.0\%$, respectively.
To reach an energy error of $10^{-4}\ \mathrm{Hartree}$, the four
approaches require 41, 55, 63 and 53 excitation operators, respectively. 
Correspondingly, HA-ADAPT-VQE reduces the ansatz size by $22.6\%$ – $34.9\%$. 

Figure~\ref{fig:3}(b) shows that \ce{H2O} displays obvious energy-plateau features
for both gradient-based (ADAPT-VQE, Pruned-ADAPT-VQE) and parameter-based
(Param-ADAPT-VQE) excitation selection schemes, as reported in refs
\cite{VQE_ADAPT_Param,VQE_ADAPT_Pruned}, which severely impedes the
optimization procedure. In contrast, HA-ADAPT-VQE sustains steady,
substantial energy reduction throughout iterations and effectively
bypasses the undesired energy plateau. To reach the chemical precision,
HA-ADAPT-VQE requires 23 excitation operators, representing a $61.7\%$
reduction compared with Param-ADAPT-VQE that needs 60 operators. In
contrast, ADAPT-VQE and Pruned-ADAPT-VQE get stuck in an energy plateau
and fail to reach the target accuracy after convergence.

Figure~\ref{fig:3}(c) shows a distinct performance difference among the four algorithms for the \ce{NH3} molecule. During iterative optimization, HA-ADAPT-VQE achieves a rapid energy decline. ADAPT-VQE and Pruned-ADAPT-VQE perform moderately, whereas Param-ADAPT-VQE yields the worst results. HA-ADAPT-VQE reaches the chemical precision with 90 excitation operators. In contrast, the other three methods fail to attain the target accuracy within the allowed ansatz size.

Besides ansatz size, measurement cost is also a critical evaluation metric for algorithms. To eliminate the influence of sampling error and achieve reliable assessment results, we define measurement cost as the total number of fermionic Hamiltonian terms whose expectation values require evaluation throughout algorithm execution. This definition is consistent with the setup of Param-ADAPT-VQE \cite{VQE_ADAPT_Param}.

Specifically, Figure~\ref{fig:3}(d) shows that HA-ADAPT-VQE
and Param-ADAPT-VQE incur higher initial measurement costs than gradient-based
ADAPT-VQE and Pruned-ADAPT-VQE, owing to the additional local parameter
optimization performed during operator pool screening. As iterations
proceed, however, HA-ADAPT-VQE and Param-ADAPT-VQE adopt the hot-starting
strategy and involve fewer redundant operators, leading to a slower
rise in measurement cost. At the energy error of $10^{-4}\ \mathrm{Hartree}$,
the measurement costs of HA-ADAPT-VQE, ADAPT-VQE, Pruned-ADAPT-VQE
and Param-ADAPT-VQE are $3.53\times10^{7}$, $7.12\times10^{7}$,
$7.33\times10^{7}$ and $6.55\times10^{7}$, respectively. Compared
with the other three algorithms, HA-ADAPT-VQE reduces the measurement
cost by $46.7\%$ - $51.9\%$. 

Figure~\ref{fig:3}(e) indicates that for the \ce{H2O}
molecule, HA-ADAPT-VQE has a measurement cost comparable to the other
three algorithms at the energy accuracy of $10^{-2}$ $\mathrm{Hartree}$.
When reaching the chemical precision, the measurement cost of HA-ADAPT-VQE
is $3.07\times10^{7}$, which represents a $76.9\%$ reduction relative
to Param-ADAPT-VQE at $1.33\times10^{8}$. ADAPT-VQE and Pruned-ADAPT-VQE
yield measurement costs of $5.51\times10^{7}$ and $5.74\times10^{7}$,
both higher than that of HA-ADAPT-VQE, yet they fail to attain the chemical
precision.

Figure~\ref{fig:3}(f) shows that Param-ADAPT-VQE
consumes the highest measurement cost of $1.28\times10^{10}$ while
achieving the lowest accuracy with an energy error of $7.6\times10^{-2}$ $\mathrm{Hartree}$.
At the energy error of $10^{-2}$ $\mathrm{Hartree}$, the measurement costs of HA-ADAPT-VQE,
ADAPT-VQE and Pruned-ADAPT-VQE are $2.74\times10^{7}$, $9.02\times10^{8}$
and $5.78\times10^{8}$, respectively. Accordingly, HA-ADAPT-VQE reduces
the measurement cost by $97.0\%$ and $95.3\%$. When reaching the chemical
precision, HA-ADAPT-VQE requires $3.09\times10^{9}$ measurements.
ADAPT-VQE and Pruned-ADAPT-VQE incur measurement costs of the same
order of magnitude, yet they only achieve an energy error of about
$9\times10^{-2}$ $\mathrm{Hartree}$.

Next, we take the \ce{H2O} molecule as an example to further explore the
underlying reasons for the performance differences among these algorithms. 

\begin{figure}
\centering
\includegraphics[width=0.37\paperwidth]{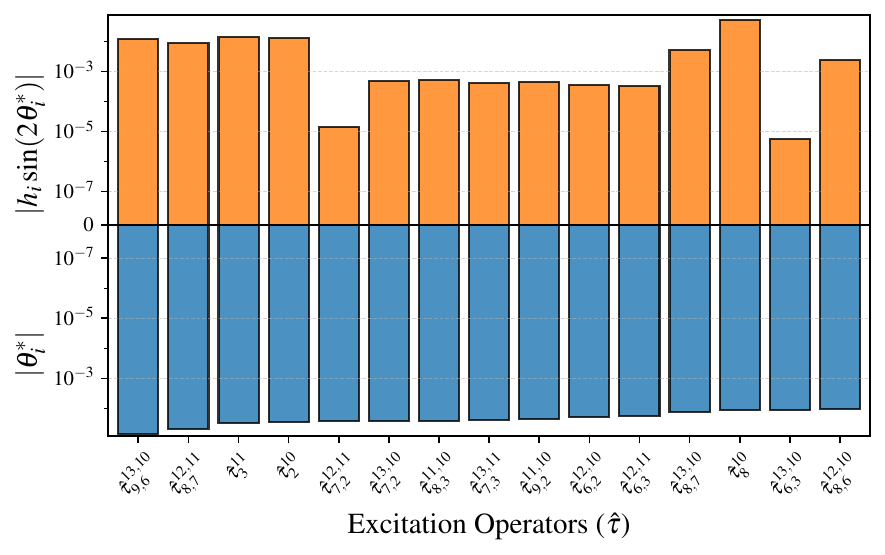}
\caption{Comparison of $|\theta_i^*|$ and weighted $|h_i\sin(2\theta_i^*)|$ for several excitation operators at the 6th iteration of HA-ADAPT-VQE (see Figure \ref{fig:3}(b)). Operators are sorted by $|\theta_i^*|$.}
\label{fig:4}
\end{figure}

Figure~\ref{fig:4} compares the magnitudes of locally optimized parameters $|\theta_i^*|$ for several excitation operators at the 6$^{\text{th}}$ iteration of HA-ADAPT-VQE (see Figure \ref{fig:3}(b)) with the corresponding values $|h_i\sin(2\theta_i^*)|$ that incorporate Hamiltonian information. The excitation operators are sorted by the magnitude of $|\theta_i^*|$. 
As observed in the figure, the $|\theta_i^*|$ values of these operators are comparable and fall into the same order of magnitude. Thus, relying solely on $|\theta_i^*|$ cannot provide a reasonable selection guideline. In contrast, $|h_i\sin(2\theta_i^*)|$ varies substantially and carries more comprehensive information. In the subsequent iteration (not plotted here), selecting the operator $\hat{\tau}_{9,6}^{13,10}$ with the largest $|\theta_i^*|$ only reduces the error by $1.01\times10^{-2}$ Hartree, while the operator $\hat{\tau}_{8}^{10}$ chosen based on $|h_i\sin(2\theta_i^*)|$ yields an energy reduction of $3.10\times10^{-2}$ Hartree. This demonstrates that the latter is a better selection criterion.

\begin{figure}
\centering
\subfloat[]{\includegraphics[width=0.37\paperwidth]{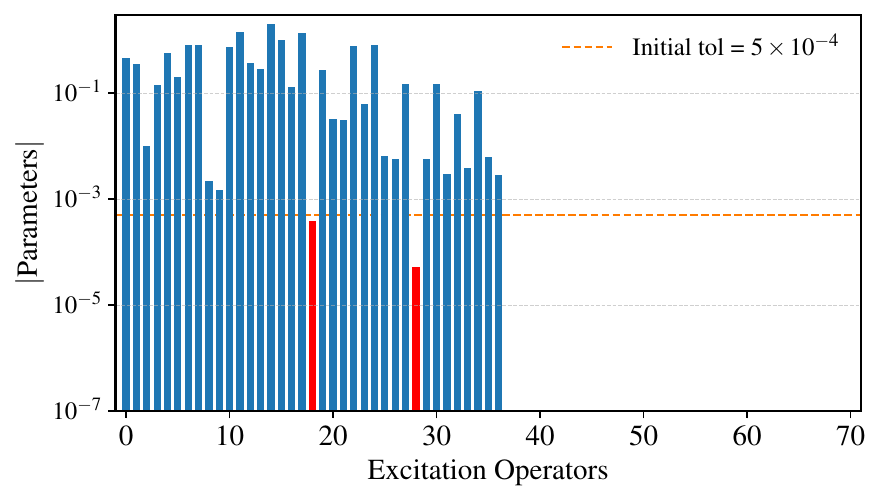}}\\
\subfloat[]{\includegraphics[width=0.37\paperwidth]{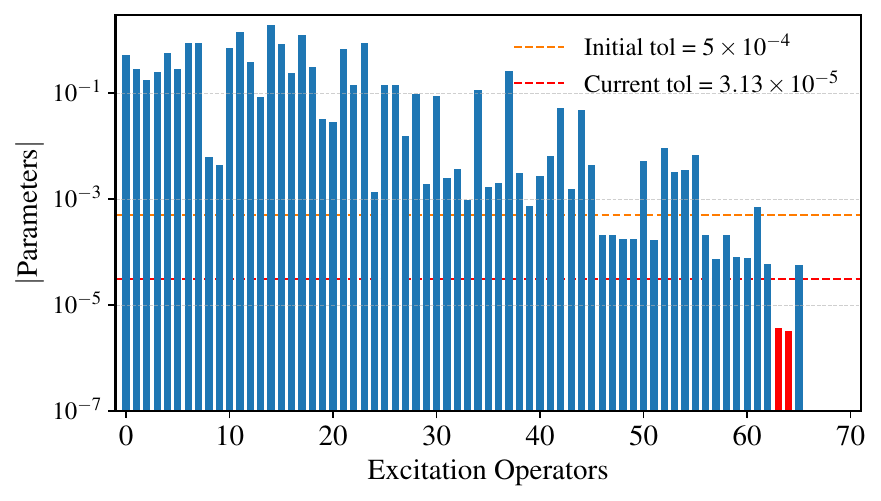}}
\caption{
For the \ce{H2O} molecule (see Figure 2(b)), as a new important excitation operator is incorporated, certain existing ones become redundant (marked in red) and are pruned by the proposed problem-adaptive pruning method. After operator removal, the energy increases by only $1.14\times10^{-9}$ and $3.00\times10^{-10}$ Hartree for case (a) and (b), respectively.
\label{fig:6}}
\end{figure}
We further analyze the effect of the dynamic threshold adopted in
HA-ADAPT-VQE. As iterations proceed and computational accuracy improves,
newly added excitation operators mainly serve for fine correction,
and their corresponding parameters gradually decrease. A fixed threshold
brings drawbacks: an excessively large threshold fails to eliminate
redundant operators at an early stage, while an overly small one continuously
discards newly added operators with small parameters and thus hinders
convergence. Therefore, a dynamically adjusted threshold that decreases
alongside iterations is essential to balance redundant operator removal
and convergence performance.

Figure~\ref{fig:6}(a) shows two excitation operators (marked in red) whose parameter values fall below the threshold $\mathrm{tol}=5\times10^{-4}$ in the ansatz. After HA-ADAPT-VQE removes these operators and re-optimizes the surviving parameters, the system energy increases by merely $1.14\times10^{-9}$ Hartree.
In most cases, removing redundant excitation operators via
HA-ADAPT-VQE leads to only a slight energy increase or even an energy
reduction. Figure~\ref{fig:6}(b) presents another
scenario. As new dominant excitation operators are incorporated, the
two small-parameter operators previously retained at the end of the
ansatz for convergence purposes are reclassified as redundant and
removed. After parameter re-optimization, the system energy increases
by only $3.00\times10^{-10}$ Hartree. These results verify the effectiveness
of the proposed removal strategy.

\begin{figure}
\centering
\subfloat[]{\includegraphics[width=0.37\paperwidth]{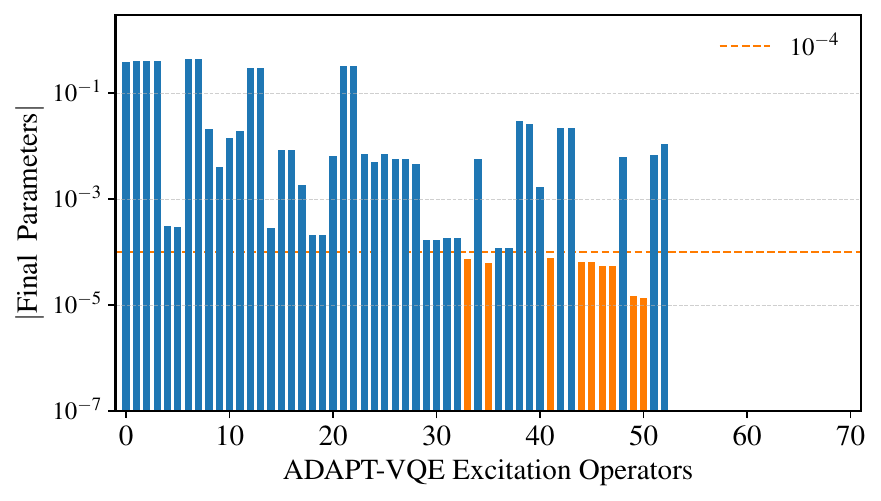}}\\
\subfloat[]{\includegraphics[width=0.37\paperwidth]{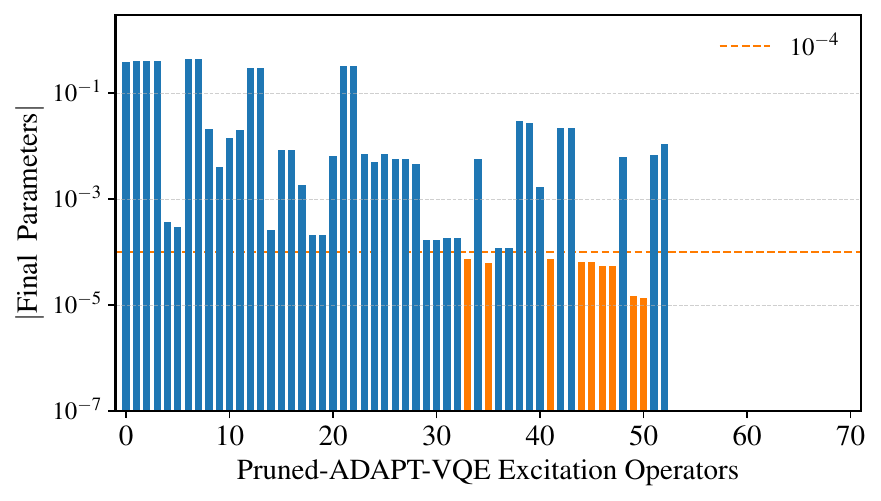}}\\
\subfloat[]{\includegraphics[width=0.37\paperwidth]{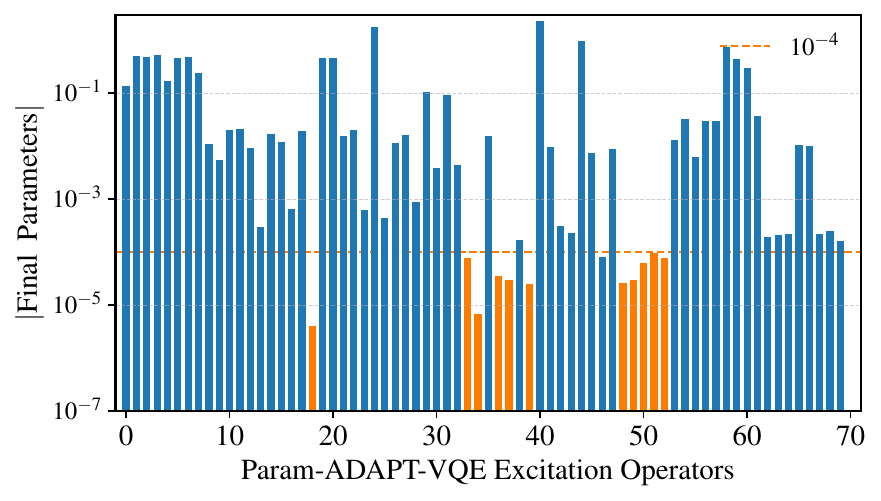}}\\
\subfloat[]{\includegraphics[width=0.37\paperwidth]{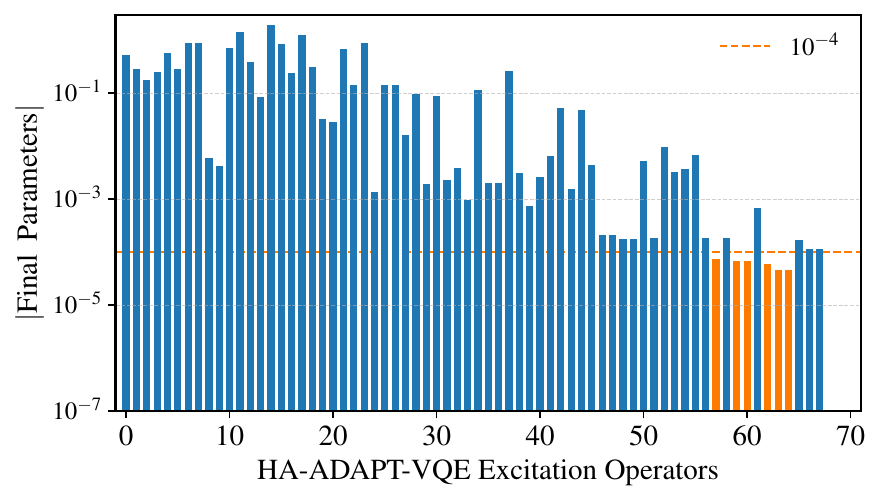}}
\caption{Comparison of final parameters across algorithms for \ce{H2O}. Small parameters (marked in orange) generally correspond to redundant excitation operators. \label{fig:5}}
\end{figure}

Figure~\ref{fig:5}(a) and (b) present the absolute values
of the final converged parameters for ADAPT-VQE and Pruned-ADAPT-VQE, respectively.
Parameters with absolute values below $10^{-4}$ are highlighted in
orange. These small parameters generally contribute negligibly to the total
energy, indicating the corresponding excitation operators are likely
redundant. It can be observed that the converged parameter values
of ADAPT-VQE and Pruned-ADAPT-VQE are nearly identical. Although Pruned-ADAPT-VQE
is designed to remove redundant excitation operators, the operators
with small parameters appear at the later positions of the ansatz.
In this case, the positional weight in the scoring function dominates
over the parameter weight, so the pruning mechanism is not activated.
Consequently, redundant excitation operators remain in the final ansatz.

Figure~\ref{fig:5}(c) and (d) illustrate the converged
parameters of Param-ADAPT-VQE and HA-ADAPT-VQE. Without a pruning
mechanism, Param-ADAPT-VQE contains numerous excitation operators
with small parameters in the front and middle parts of the ansatz,
leading to high redundancy. For HA-ADAPT-VQE, such small-parameter
operators only appear at the tail of the ansatz to guarantee convergence.
This demonstrates that the pruning strategy of HA-ADAPT-VQE effectively
eliminates redundant operators across all positions while maintaining
good convergence performance.

\section{CONCLUSION }\label{sec:4}

In this work, we present the HA-ADAPT-VQE algorithm, which combines a novel excitation operator selection criterion and a problem-adaptive scheme to prune redundant operators.  Traditional ADAPT-VQE relies solely on local information of candidate excitation operators such as energy gradients. By contrast, our new criterion incorporates locally optimized parameters and corresponding Hamiltonian coefficients. It effectively captures intrinsic orbital interaction strength, enabling the algorithm to prioritize physically meaningful operators and suppress ineffective terms with negligible Hamiltonian contributions.
Our pruning method assesses all operators within the ansatz according to their parameters and positions, and divides them into three categories to indicate whether they should be retained, removed or tentatively kept to ensure convergence. According to the pruning results, it can automatically adjust the parameter thresholds to guarantee convergence, making it applicable to ansätze of any scale.
We perform experiments on three representative strongly correlated molecules: stretched \ce{BeH2}, \ce{H2O}, and \ce{NH3}. The results show that HA-ADAPT-VQE outperforms ADAPT-VQE, Pruned-ADAPT-VQE and Param-ADAPT-VQE in energy error, ansatz size and measurement cost. This work offers a useful reference and helps advance both experimental and theoretical research on large-scale VQE algorithms for quantum chemistry.

\section*{ACKNOWLEDGMENTS}
This work was supported by the Innovation Program for Quantum Science and Technology (Grant Nos.~2024ZD0300502, 2024ZD0300500), the Beijing Nova Program (Grant No. 20240484652), the Youth Innovation Promotion Association of the Chinese Academy of Sciences (Grant No. 2023116), the National Natural Science Foundation of China (Grant No. 62402485), the Young Elite Scientists Sponsorship Program of the China Association for Science and Technology, the CCF-QuantumCtek Special Cooperation Program on Superconducting Quantum Computing (Grant No. CCF-QC2025007), the International Partnership Program of the Chinese Academy of Sciences (Grant No. 096GJHZ2025013FN), and the CPS-Yangtze Delta Region Industrial Innovation Center of Quantum and Information Technology-MindSpore Quantum Open Fund.

\bibliographystyle{unsrt}
\bibliography{References_library}

\begin{thebibliography}{10}

\bibitem{molecular_property_calculations}
Trygve Helgaker, Sonia Coriani, Poul J{\o}rgensen, Kasper Kristensen, Jeppe Olsen, and Kenneth Ruud.
\newblock Recent advances in wave function-based methods of molecular-property calculations.
\newblock {\em Chemical reviews}, 112(1):543--631, 2012.

\bibitem{Eyring_equation}
Henry Eyring.
\newblock The activated complex in chemical reactions.
\newblock {\em The Journal of chemical physics}, 3(2):107--115, 1935.

\bibitem{szabo}
Attila Szabo and Neil~S Ostlund.
\newblock {\em Modern quantum chemistry: introduction to advanced electronic structure theory}.
\newblock Courier Corporation, 1996.

\bibitem{CCSD}
George~D Purvis~III and Rodney~J Bartlett.
\newblock A full coupled-cluster singles and doubles model: The inclusion of disconnected triples.
\newblock {\em The Journal of chemical physics}, 76(4):1910--1918, 1982.

\bibitem{DFT}
Min-Cheol Kim, Eunji Sim, and Kieron Burke.
\newblock Understanding and reducing errors in density functional calculations.
\newblock {\em Physical review letters}, 111(7):073003, 2013.

\bibitem{DFT_1}
Walter Kohn and Lu~Jeu Sham.
\newblock Self-consistent equations including exchange and correlation effects.
\newblock {\em Physical review}, 140(4A):A1133, 1965.

\bibitem{strongly_correlated_systems}
Zhendong Li, Sheng Guo, Qiming Sun, and Garnet Kin-Lic Chan.
\newblock Electronic landscape of the p-cluster of nitrogenase as revealed through many-electron quantum wavefunction simulations.
\newblock {\em Nature chemistry}, 11(11):1026--1033, 2019.

\bibitem{CCSD_fail_1}
J~Paldus, J~{\v{C}}{\'\i}{\v{z}}ek, and M~Takahashi.
\newblock Approximate account of the connected quadruply excited clusters in the coupled-pair many-electron theory.
\newblock {\em Physical Review A}, 30(5):2193, 1984.

\bibitem{CCSD_fail_2}
David~W Small and Martin Head-Gordon.
\newblock A fusion of the closed-shell coupled cluster singles and doubles method and valence-bond theory for bond breaking.
\newblock {\em The Journal of chemical physics}, 137(11), 2012.

\bibitem{qc_nielsen}
Michael~A Nielsen and Isaac~L Chuang.
\newblock {\em Quantum Computation and Quantum Information 10th Anniversary Edition}.
\newblock Cambridge University Press, 2010.

\bibitem{QPE}
Daniel~S Abrams and Seth Lloyd.
\newblock Quantum algorithm providing exponential speed increase for finding eigenvalues and eigenvectors.
\newblock {\em Physical Review Letters}, 83(24):5162, 1999.

\bibitem{NISQ}
John Preskill.
\newblock Quantum computing in the nisq era and beyond.
\newblock {\em Quantum}, 2:79, 2018.

\bibitem{NISQ_1}
Frank Leymann and Johanna Barzen.
\newblock The bitter truth about gate-based quantum algorithms in the nisq era.
\newblock {\em Quantum Science and Technology}, 5(4):044007, 2020.

\bibitem{VQE_first}
Alberto Peruzzo, Jarrod McClean, Peter Shadbolt, Man-Hong Yung, Xiao-Qi Zhou, Peter~J Love, Al{\'a}n Aspuru-Guzik, and Jeremy~L O’brien.
\newblock A variational eigenvalue solver on a photonic quantum processor.
\newblock {\em Nature communications}, 5(1):4213, 2014.

\bibitem{VQE_Science}
Google~AI Quantum, Collaborators, Frank Arute, Kunal Arya, Ryan Babbush, Dave Bacon, Joseph~C Bardin, Rami Barends, Sergio Boixo, Michael Broughton, Bob~B Buckley, et~al.
\newblock Hartree-fock on a superconducting qubit quantum computer.
\newblock {\em Science}, 369(6507):1084--1089, 2020.

\bibitem{VQE_review_0}
Sam McArdle, Suguru Endo, Alan Aspuru-Guzik, Simon~C Benjamin, and Xiao Yuan.
\newblock Quantum computational chemistry.
\newblock {\em Reviews of Modern Physics}, 92(1):015003, 2020.

\bibitem{VQE_review_1}
Jules Tilly, Hongxiang Chen, Shuxiang Cao, Dario Picozzi, Kanav Setia, Ying Li, Edward Grant, Leonard Wossnig, Ivan Rungger, George~H Booth, et~al.
\newblock The variational quantum eigensolver: a review of methods and best practices.
\newblock {\em Physics Reports}, 986:1--128, 2022.

\bibitem{VQE_review_2}
Marco Cerezo, Andrew Arrasmith, Ryan Babbush, Simon~C Benjamin, Suguru Endo, Keisuke Fujii, Jarrod~R McClean, Kosuke Mitarai, Xiao Yuan, Lukasz Cincio, et~al.
\newblock Variational quantum algorithms.
\newblock {\em Nature Reviews Physics}, 3(9):625--644, 2021.

\bibitem{VQE_ucc_review}
Jonathan Romero, Ryan Babbush, Jarrod~R McClean, Cornelius Hempel, Peter~J Love, and Al{\'a}n Aspuru-Guzik.
\newblock Strategies for quantum computing molecular energies using the unitary coupled cluster ansatz.
\newblock {\em Quantum Science and Technology}, 4(1):014008, 2018.

\bibitem{VQE_HEA_nature}
Abhinav Kandala, Antonio Mezzacapo, Kristan Temme, Maika Takita, Markus Brink, Jerry~M Chow, and Jay~M Gambetta.
\newblock Hardware-efficient variational quantum eigensolver for small molecules and quantum magnets.
\newblock {\em Nature}, 549(7671):242--246, 2017.

\bibitem{Barren_plateaus}
Jarrod~R McClean, Sergio Boixo, Vadim~N Smelyanskiy, Ryan Babbush, and Hartmut Neven.
\newblock Barren plateaus in quantum neural network training landscapes.
\newblock {\em Nature communications}, 9(1):4812, 2018.

\bibitem{VQE_ADAPT}
Harper~R Grimsley, Sophia~E Economou, Edwin Barnes, and Nicholas~J Mayhall.
\newblock An adaptive variational algorithm for exact molecular simulations on a quantum computer.
\newblock {\em Nature communications}, 10(1):3007, 2019.

\bibitem{VQE_ADAPT_application}
Christian~W Bauer, Zohreh Davoudi, Natalie Klco, and Martin~J Savage.
\newblock Quantum simulation of fundamental particles and forces.
\newblock {\em Nature Reviews Physics}, 5(7):420--432, 2023.

\bibitem{VQE_ADAPT_measurement_reuse}
Azhar Ikhtiarudin, Gagus~Ketut Sunnardianto, Fadjar Fathurrahman, Mohammad~Kemal Agusta, and Hermawan~Kresno Dipojono.
\newblock Shot-efficient adapt-vqe via reused pauli measurements and variance-based shot allocation.
\newblock {\em arXiv preprint arXiv:2507.16879}, 2025.

\bibitem{VQE_ADAPT_pool_simutameously_measurement}
Panagiotis~G Anastasiou, Nicholas~J Mayhall, Edwin Barnes, and Sophia~E Economou.
\newblock How to really measure operator gradients in adapt-vqe.
\newblock {\em arXiv preprint arXiv:2306.03227}, 2023.

\bibitem{VQE_ADAPT_qubit}
Ho~Lun Tang, VO~Shkolnikov, George~S Barron, Harper~R Grimsley, Nicholas~J Mayhall, Edwin Barnes, and Sophia~E Economou.
\newblock qubit-adapt-vqe: An adaptive algorithm for constructing hardware-efficient ans{\"a}tze on a quantum processor.
\newblock {\em PRX Quantum}, 2(2):020310, 2021.

\bibitem{VQE_ADAPT_TETRIS}
Panagiotis~G Anastasiou, Yanzhu Chen, Nicholas~J Mayhall, Edwin Barnes, and Sophia~E Economou.
\newblock Tetris-adapt-vqe: An adaptive algorithm that yields shallower, denser circuit ans{\"a}tze.
\newblock {\em Physical Review Research}, 6(1):013254, 2024.

\bibitem{VQE_ADAPT_amp_reording_batch}
Zhihao Lan and WanZhen Liang.
\newblock Amplitude reordering accelerates the adaptive variational quantum eigensolver algorithms.
\newblock {\em Journal of Chemical Theory and Computation}, 18(9):5267--5275, 2022.

\bibitem{VQE_ADAPT_CEO}
Mafalda Ram{\^o}a, Panagiotis~G Anastasiou, Luis~Paulo Santos, Nicholas~J Mayhall, Edwin Barnes, and Sophia~E Economou.
\newblock Reducing the resources required by adapt-vqe using coupled exchange operators and improved subroutines.
\newblock {\em npj Quantum Information}, 11(1):86, 2025.

\bibitem{VQE_ADAPT_reduced_RDM}
Jie Liu, Zhenyu Li, and Jinlong Yang.
\newblock An efficient adaptive variational quantum solver of the schr{\"o}dinger equation based on reduced density matrices.
\newblock {\em The Journal of chemical physics}, 154(24), 2021.

\bibitem{VQE_ADAPT_Param}
Runhong He, Xin Hong, Qiaozhen Chai, Ji~Guan, Junyuan Zhou, Arapat Ablimit, Guolong Cui, and Shenggang Ying.
\newblock Constructing compact adapt unitary coupled-cluster ansatz with parameter-based criterion.
\newblock {\em Journal of Chemical Theory and Computation}, 22(10):5090--5101, 2026.
\newblock PMID: 42127224.

\bibitem{VQE_ADAPT_QEB}
Yordan~S Yordanov, Vasileios Armaos, Crispin~HW Barnes, and David~RM Arvidsson-Shukur.
\newblock Qubit-excitation-based adaptive variational quantum eigensolver.
\newblock {\em Communications Physics}, 4(1):228, 2021.

\bibitem{VQE_ADAPT_ES}
Yi~Fan, Changsu Cao, Xusheng Xu, Zhenyu Li, Dingshun Lv, and Man-Hong Yung.
\newblock Circuit-depth reduction of unitary-coupled-cluster ansatz by energy sorting.
\newblock {\em The Journal of Physical Chemistry Letters}, 14(43):9596--9603, 2023.

\bibitem{VQE_ADAPT_Pruned}
Nonia Vaquero-Sabater, Abel Carreras, and David Casanova.
\newblock Pruned-adapt-vqe: compacting molecular ansatze by removing irrelevant operators.
\newblock {\em Journal of Chemical Theory and Computation}, 21(18):8720--8728, 2025.

\bibitem{VQE_HiUCCSD}
Runhong He, Arapat Ablimit, Xin Hong, Qiaozhen Chai, Junyuan Zhou, Ji~Guan, Guolong Cui, and Shenggang Ying.
\newblock Hamiltonian-informed point group symmetry-respecting ansätze for the variational quantum eigensolver.
\newblock {\em Journal of Chemical Theory and Computation}, 0(0):null, 0.

\bibitem{RRvariational}
Sydney~Henry Gould.
\newblock {\em Variational methods for eigenvalue problems: an introduction to the Weinstein method of intermediate problems}.
\newblock University of Toronto Press, 1966.

\bibitem{VQE_trotter_one}
Panagiotis~Kl Barkoutsos, Jerome~F Gonthier, Igor Sokolov, Nikolaj Moll, Gian Salis, Andreas Fuhrer, Marc Ganzhorn, Daniel~J Egger, Matthias Troyer, Antonio Mezzacapo, et~al.
\newblock Quantum algorithms for electronic structure calculations: Particle-hole hamiltonian and optimized wave-function expansions.
\newblock {\em Physical Review A}, 98(2):022322, 2018.

\bibitem{Jordan_Wigner}
Pascual Jordan and Eugene~Paul Wigner.
\newblock {\em {\"U}ber das paulische {\"a}quivalenzverbot}.
\newblock Springer, 1993.

\bibitem{VQE_fops_9_cnot}
Zhijie Sun, Xiaopeng Li, Jie Liu, Zhenyu Li, and Jinlong Yang.
\newblock Circuit-efficient qubit excitation-based variational quantum eigensolver.
\newblock {\em Journal of Chemical Theory and Computation}, 2025.

\bibitem{VQE_fops_13_cnot}
Yordan~S Yordanov, David~RM Arvidsson-Shukur, and Crispin~HW Barnes.
\newblock Efficient quantum circuits for quantum computational chemistry.
\newblock {\em Physical Review A}, 102(6):062612, 2020.

\bibitem{VQE_SymUCCSD}
Changsu Cao, Jiaqi Hu, Wengang Zhang, Xusheng Xu, Dechin Chen, Fan Yu, Jun Li, Han-Shi Hu, Dingshun Lv, and Man-Hong Yung.
\newblock Progress toward larger molecular simulation on a quantum computer: Simulating a system with up to 28 qubits accelerated by point-group symmetry.
\newblock {\em Physical Review A}, 105(6):062452, 2022.

\bibitem{VQE_SymUCCSD_failed}
Leon~D da~Silva and Marcelo~P Santos.
\newblock Lie-algebraic incompleteness of symmetry-adapted vqe for non-abelian molecular point groups.
\newblock {\em arXiv preprint arXiv:2603.21009}, 2026.

\bibitem{VQE_ADAPT_vs_bp}
Harper~R Grimsley, George~S Barron, Edwin Barnes, Sophia~E Economou, and Nicholas~J Mayhall.
\newblock Adaptive, problem-tailored variational quantum eigensolver mitigates rough parameter landscapes and barren plateaus.
\newblock {\em npj Quantum Information}, 9(1):19, 2023.

\bibitem{VQE_ADAPT_grad_free}
Jonas J{\"a}ger, Thierry~N Kaldenbach, Max Haas, and Erik Schultheis.
\newblock Fast gradient-free optimization of excitations in variational quantum eigensolvers.
\newblock {\em Communications Physics}, 8(1):418, 2025.

\bibitem{MindQuantum}
Xusheng Xu, Jiangyu Cui, Zidong Cui, Runhong He, Qingyu Li, Xiaowei Li, Yanling Lin, Jiale Liu, Wuxin Liu, Jiale Lu, Maolin Luo, Chufan Lyu, Shijie Pan, Mosharev Pavel, Runqiu Shu, Jialiang Tang, Ruoqian Xu, Shu Xu, Kang Yang, Fan Yu, Qingguo Zeng, Haiying Zhao, Qiang Zheng, Junyuan Zhou, Xu~Zhou, Yikang Zhu, Zuoheng Zou, Abolfazl Bayat, Xi~Cao, Wei Cui, Zhendong Li, Guilu Long, Zhaofeng Su, Xiaoting Wang, Zizhu Wang, Shijie Wei, Re-Bing Wu, Pan Zhang, and Man-Hong Yung.
\newblock Mindspore quantum: A user-friendly, high-performance, and ai-compatible quantum computing framework, 2024.

\bibitem{parameter_shift_rule_0}
Maria Schuld, Ville Bergholm, Christian Gogolin, Josh Izaac, and Nathan Killoran.
\newblock Evaluating analytic gradients on quantum hardware.
\newblock {\em Physical Review A}, 99(3):032331, 2019.

\bibitem{Parameter_shift}
Ryan Sweke, Frederik Wilde, Johannes Meyer, Maria Schuld, Paul~K F{\"a}hrmann, Barth{\'e}l{\'e}my Meynard-Piganeau, and Jens Eisert.
\newblock Stochastic gradient descent for hybrid quantum-classical optimization.
\newblock {\em Quantum}, 4:314, 2020.

\bibitem{bfgs}
Charles~G Broyden.
\newblock Quasi-newton methods and their application to function minimisation.
\newblock {\em Mathematics of Computation}, 21(99):368--381, 1967.

\bibitem{pyscf}
Qiming Sun, Timothy~C Berkelbach, Nick~S Blunt, George~H Booth, Sheng Guo, Zhendong Li, Junzi Liu, James~D McClain, Elvira~R Sayfutyarova, Sandeep Sharma, et~al.
\newblock Pyscf: the python-based simulations of chemistry framework.
\newblock {\em Wiley Interdisciplinary Reviews: Computational Molecular Science}, 8(1):e1340, 2018.

\bibitem{scipy}
Pauli Virtanen, Ralf Gommers, Travis~E Oliphant, Matt Haberland, Tyler Reddy, David Cournapeau, Evgeni Burovski, Pearu Peterson, Warren Weckesser, Jonathan Bright, et~al.
\newblock Scipy 1.0: fundamental algorithms for scientific computing in python.
\newblock {\em Nature methods}, 17(3):261--272, 2020.

\end{thebibliography}
\end{document}